\documentstyle[epsf,prd,aps,floats]{revtex}
\begin{document}
\twocolumn[\hsize\textwidth\columnwidth\hsize\csname@twocolumnfalse\endcsname

\title{Could thermal fluctuations seed cosmic structure?}

\author{
Jo\~ao Magueijo and Levon Pogosian}
\address{
Theoretical Physics, The Blackett Laboratory, Imperial College,\\
Prince Consort Road, London SW7 2BZ, United Kingdom }

\maketitle

\begin{abstract}
We examine the possibility that thermal, rather than
quantum, fluctuations are responsible for seeding the structure of our universe.
We find that while the thermalization condition leads to
nearly Gaussian statistics,
a Harrisson-Zeldovich spectrum for the primordial fluctuations can only be achieved
in very special circumstances.
These depend on whether the universe gets hotter or colder in time, while
the modes are leaving the horizon. In the latter case we find a no-go theorem
which can only be avoided if the fundamental degrees of freedom are
not particle-like, such as in string gases near the Hagedorn phase
transition. The former case is less forbidding, and we suggest two
potentially successful ``warming universe'' scenarios. One makes use of the
Phoenix universe, the other of ``phantom'' matter.
\end{abstract}
]

\renewcommand{\thefootnote}{\arabic{footnote}}
\setcounter{footnote}{0}



\section{Introduction}
A major success of the Big Bang theory is its ability to account
for the detailed structure of the Universe, such as galaxy
clustering, or the temperature fluctuations in the cosmic
microwave background. Paramount to this picture is the phenomenon
of gravitational instability, by means of which primordial small
departures from homogeneity can grow into the observed structures.
Within the Big Bang theory the required primordial fluctuations
are treated merely as initial conditions. Vacuum {\it quantum} fluctuations
in inflationary scenarios have been shown to lead to the required initial conditions,
fitting current large scale structure data. It is, however, important to know just
how unavoidable is the conclusion that cosmic structures have a quantum origin.
Could these primordial fluctuations have a {\it thermal} origin instead?

The possibility that we are descendent from thermal fluctuations was
advanced by Peebles, in  his book {\it Principles of Physical
Cosmology}, pp.~371-373  \cite{peebles}. Peebles pointed out that if the
Universe was in thermal equilibrium on the comoving scale of 10
Mpc when its temperature was $T=10^{11}$~Gev, then the observed
value of $\sigma_{10}$ could be explained. This fascinating remark leaves
several questions unanswered. Such a scenario may explain the observed value
of $\sigma_{10}$, but what about fluctuations on other scales? Furthermore, thermal
fluctuations are not strictly Gaussian - does this scenario
conflict with observations?

Motivated by these unsolved issues, in this paper we go further
and examine under which conditions primordial thermal fluctuations
lead to a Harrison-Zeldovich (HZ) \cite{hz} spectrum of
approximately Gaussian fluctuations. Such a scenario would fit
{\it all} existing data in the same way that the usual
inflationary quantum fluctuation scenario does. Indeed, the only
potentially distinguishing feature would be a different signature
(or absence thereof) of gravitational waves (tensor modes) in the
thermal scenario.

A matter not addressed by Peebles is how to establish
thermalization on the relevant scales. In pure Big Bang cosmology
all observed fluctuations spanned causally disconnected
regions when the universe was at the required high temperatures to
induce the appropriate level of fluctuations. In addition, simply
{\it postulating} exact thermalization over all scales, say at Planck time,
leads to grossly inappropriate results (thermal fluctuations are
white-noise rather than scale invariant). Hence, we should assume
that exact thermalization only applies to sub-horizon modes,
and that there is a mechanism for pushing sub-horizon thermal modes
outside the horizon, where they freeze and become non-thermal. Since different
modes freeze at different temperatures, it may be that the final
super-horizon spectrum is indeed of HZ type.

We shall consider three types of mechanisms for pushing modes
outside the horizon:  accelerated expansion, varying speed of
light (VSL) and a contracting universe. In the first case we
consider inflation models driven by a dominant component of
thermal radiation, as in \cite{ncinfl}, where deformed dispersion
relations affect common radiation at high temperatures. This is
not to be confused with inflationary models in which there is a
finite radiation component during inflation
\cite{yokoyama,thermalinf}. There the dominant component is always
the inflaton field (even though the thermal bath drives inflaton
fluctuations). The second possibility is a varying speed of light
(VSL), either in the form of a space-time field ($c(x^\mu)$)
\cite{moffat93,am,mofclay,covvsl}, or as an energy-dependent
effect ($c(E)$) ~\cite{ncvsl}. There are VSL models \cite{mofclay}
in which quantum vacuum fluctuations can produce the HZ
spectrum\cite{moffat}. However, in a large class of VSL scenarios
the universe is never vacuum dominated. Hence subhorizon scale
thermalization is the reason for the apparent homogeneity of the
Universe - and likewise thermal fluctuations are responsible for
the primordial fluctuations. In the third case, we consider the
possibility of thermal fluctuations seeding structures within
Lemaitre's Phoenix universe \cite{lem}.

The paper is organized as follows. In Section \ref{gaussianity} we examine
the statistics of thermal fluctuations. In Section \ref{powerspectrum} we
derive the necessary conditions for thermal fluctuations to have a scale-invariant
spectrum. In Section \ref{nogo} we show that these conditions cannot be fulfilled
in universes dominated by radiation comprised of conventional particles and in which
the temperature decreases with time. At the end of Section \ref{nogo} and
in Section \ref{warming} we propose a set of models that may bypass this
no-go theorem. We summarize our results in Section \ref{summary}.

\section{The statistics of thermal fluctuations}
\label{gaussianity}

Most of the currently available data is
consistent with the hypothesis that primordial fluctuations are Gaussian distributed
\cite{gauss}. A possible exception is the report of non-vanishing ``inter-scale''
components of the CMB bispectrum calculated from the four year
COBE-DMR data \cite{nongauss}.
New CMB measurements, in particular by the MAP satellite, are expected to provide
tighter constraints on the amount of cosmological non-Gaussianity.

Based on current observational evidence, one has to require that at least on very
large scales, primordial fluctuations must be sufficiently well described by a
Gaussian distribution. Here we encounter the first obstacle, since strictly speaking
thermal fluctuations are not Gaussian. In what follows we shall show that these
fluctuations are Gaussian to a very good approximation under the same set of
conditions which assure thermalization.

Fluctuations in a thermal (canonical) ensemble can be determined
from the partition function
\begin{equation}
Z={\sum_r} e^{-\beta E_r} \ ,
\end{equation}
where $\beta = T^{-1}$.
This is true even if deformed dispersion relations are introduced
\cite{ncvsl}. Hence the total energy $U$ inside a volume $V$ is
given by:
\begin{equation}
U={\langle E\rangle}={{\sum _r} E_r e^{-\beta E_r} \over {\sum_r}
e^{-\beta E_r}}=-{d\log Z\over d\beta}
\end{equation}
In general this integral needs not be proportional to $T^4$, and
indeed under deformed dispersion relations $U\propto T^\gamma$,
with $1<\gamma<4$. However, because energy is an extensive
quantity, we always have that $U=\rho V$, that is $U$ is
proportional to the volume. If $\gamma\neq 4$ there is a preferred
length scale $l_T$, and we can choose units so that
\begin{equation}\label{rhoeq}
\rho\approx T^4 (l_T T)^{\gamma -4}.
\end{equation}
(specifically we use units such that $G=\hbar=c_0=k_B=1$, where $c_0$ is current
value of $c$, and neglect factors of order 1).

The energy variance is given by
\begin{equation}
\sigma^2(E)={\langle E^2\rangle}-{\langle E\rangle}^2={d^2\log
Z\over d\beta^2}= -{dU \over d\beta}
\end{equation}
and so the relative variance is
\begin{equation}\label{var}
\sigma^2_U={\sigma^2(E)\over U^2}={T^2\rho'\over \rho^2}{1\over V}
\end{equation}
where prime denotes differentiation with respect to temperature. We
see that whereas the amplitude of these fluctuations is model
dependent, their spectrum (i.e. the fact that $\sigma^2_U\propto
1/ V$) is not. The white noise nature of thermal fluctuations
follows from the fact that energy is an extensive quantity
(i.e. proportional to the volume).

The fact that $\sigma^2_U\propto 1/ V$ leads to an interesting
heuristic interpretation of thermal fluctuations. It seems to imply that
thermal fluctuations may be seen as a Poisson process involving a
set of regions with coherence length $\lambda$ dependent only on
the temperature (and not the sample volume). In a volume $V$ there
are $n=V/\lambda^3$ such regions, so that a Poisson process
results in variance
\begin{equation}
\sigma^2_U={\sigma^2(n)\over
{\overline n}^2}={1\over \overline{n}} ={\lambda^3\over V}
\end{equation}
implying that any white noise spectrum of fluctuations may be seen
as a Poisson process. The dependence $\lambda(T)$ can be inferred
from eq.~(\ref{var}) and in general takes the form
\begin{equation}
\lambda^3= {T^2\rho'\over \rho^2}
\end{equation}
translating, for (\ref{rhoeq}), into
\begin{equation}\label{corleng}
\lambda^3\approx {\gamma\over T^3}{1\over (l_TT)^{\gamma-4}}
\end{equation}
We see that $\lambda\propto T^{-1}$ only for $\gamma=4$. For
$\gamma=1$ (realized in non-commutative geometry \cite{ncvsl}) $\lambda$ is
temperature independent and equals the length scale of
non-commutativity. In general the thermal coherence length
decreases with increasing temperature, {\it except} if $\gamma<1$.
In the exceptional case $\gamma<1$ the coherence length increases
with the temperature and the relative energy fluctuations anomalously
increase with the temperature; in Section~\ref{nogo} we shall
rule out this exceptional case.

The non-Gaussianity of thermal fluctuations may now be studied in
terms of the cumulants of the distribution. The third order
centered cumulant is given by \cite{KendalStuart}
\begin{equation}
\kappa_3={\langle E^3\rangle}-3{\langle E^2\rangle}{\langle E
\rangle}+ 2{\langle E\rangle}^3=-{d^3\log Z\over d\beta^3}=
{d^2U\over d\beta^2}
\end{equation}
so that the relative skewness is
\begin{equation}
s_3={\kappa_3\over\sigma^3}\approx{\gamma(\gamma+1)
\over \gamma^{3/2}}
{(l_T T)^{2-{\gamma\over 2}}\over (V T^3)^{1/2}}
\end{equation}
Hence for large volumes $s_3\ll 1$.
Likewise for higher cumulants
\begin{equation}
\kappa_n=(-1)^n{d^n\log Z\over d\beta^n}= (-1)^{n-1}{d^{n-1}U\over
d\beta^{n-1}}
\end{equation}
and so
\begin{equation}
s_n={\kappa_n\over\sigma^n}\approx c_n(\gamma) {(l_T T)^{({n\over 2}-1)(4-\gamma)}
\over (VT^3)^{{n\over 2}-1}}
\end{equation}
with the proportionality constant $c_n(\gamma)=\gamma(\gamma+1)...
(\gamma+n-2)/\gamma^{n/2}$. Setting $V=L^3$ we find the
unsurprising result that this can be written as
\begin{equation}\label{cumul}
s_n\approx {c_n(\gamma)\over \gamma ^{{n\over 2}-1}}
{\left(\lambda\over L\right)}^{3({n\over 2}-1)}
\end{equation}
and so if the volume under study is much larger than the thermal
coherence length as defined above we have indeed that $s_n\ll 1$.

We conclude that thermal fluctuations are
Gaussian to a very good approximation if, when they leave the horizon,
the thermal coherence length is much smaller than the horizon size. The
departures from Gaussianity always lead to positive cumulants and
these decay with the order $n$ as in eq.~(\ref{cumul}). Since the condition
for thermalization is that the coherence length is much smaller
than the scales under study, we may conclude that Gaussianity is
part and parcel of the self-consistency conditions for studying
thermal fluctuations in thermal equilibrium.

\section{The power spectrum}
\label{powerspectrum}

Gaussian fluctuations are fully described by their two-point function.
This can be encoded in $\sigma_U^2$, but is more often expressed
in terms of the power spectrum $P(k)={\langle |\delta_k|^2\rangle}$,
where $\delta_k$ are the Fourier modes of the density contrast
$\delta\rho/\rho$.
The two can be related via the integral \cite{luccin}:
\begin{equation}
\sigma_U^2 = {1 \over 2\pi^2} \int_0^{\infty} P(k) W_F^2(kL) k^2 dk \ ,
\label{variance3}
\end{equation}
where $W_F(kL)$ is a filter function and
$L \sim V^{1/3}$ is the smoothing scale (so that $U=\rho V$).
Assuming a power law dependence for the power spectrum,
$P(k)=A^2 k^n$, and, for instance, a Gaussian filter function, $W_F(kL)=e^{-k^2L^2/2}$,
integration of eq.~(\ref{variance3}) gives
\begin{equation}
\sigma_U^2 = {\tilde{A}^2 \over L^{3+n}} ={\tilde{A}^2 \over V^{1+{n\over 3}}} \ ,
\label{variance4}
\end{equation}
where $\tilde{A}^2 = A^2 \Gamma\left( {n+3 \over 2} \right) /( 4\pi^2)$, that is,
$\tilde{A} \approx A$.
By comparing eqns.~(\ref{var}) and
(\ref{variance4}) one can see that thermal
density fluctuations have a white noise spectrum ($n=0$)
with amplitude:
\begin{equation}
\label{amplT}
P(k)={\langle {|\delta_k|^2} \rangle} \approx {T^2\rho'\over \rho^2} k^0
\end{equation}
where we have ignored factors of order 1. This result
only applies to modes which are in causal contact, i.e. sub-horizon modes.
More precisely it only applies when self-gravity
is negligible, and so to modes smaller than the
Jeans length.

Suppose we have a model in which a certain range of Fourier modes
of thermal density fluctuations are forced outside the horizon and
are allowed to re-enter it at a later time\footnote{This condition
is more restrictive than requiring that the horizon problem be
solved -- for instance the Milne universe (with $a\propto t$) does
not have horizons, and yet sub-horizon modes are not pushed
outside the horizon. The reader is also referred to
Ref.~\cite{cosmic} for a simple condition for solving the horizon
and flatness problems.}. Then, at first horizon crossing we have
\begin{equation}
\label{firstcrossing}
k_h \sim {|\dot{a}|\over c}
\end{equation}
where $k$ are comoving wavenumbers, $a$ is the scale factor and $c$ is the speed of
light\footnote{$c$ is the speed of light at the time of the horizon crossing, which could
be different from $c_0$}. The modulus sign in eq. (\ref{firstcrossing}) accounts
for the possibility of such a crossing happening during a
contracting phase in a bouncing universe - a model
discussed in subsection \ref{bouncing}. We parameterize the temperature
dependence of the first horizon crossing with
\begin{equation}\label{khor}
k_h=T^\mu \lambda_1^{\mu-1} \ ,
\end{equation}
where $\lambda_1$ is some characteristic length scale.
We will consider two types of scenarios: those in which the temperature
of the radiation is decreasing with the evolution, and those in which the
temperature is increasing.

If the universe cools as it expands then solving the horizon problem and
the existence of a first crossing require that $\mu<0$. More
generally, one needs $dk_h/dT<0$ or, from eq.~(\ref{khor}),
\begin{equation}
{\ddot a\over \dot a}-{\dot c\over c}>0
\end{equation}
Hence, while modes are being forced outside the horizon one must
have either accelerated expansion or a decreasing speed of light,
or a combination of both.

If the universe heats up as it evolves in time, then $\mu>0$ (or, $dk_h/dT>0$).
We consider this case in some detail in Section \ref{warming}.

In both types of scenarios we will be interested in identifying
conditions for a HZ spectrum of
density fluctuations to be left outside the horizon. If there is no
significant evolution of the gravitational potential $\phi$
outside the horizon while the
modes are being pushed out (this is usually enforced by requiring
that the equation of state remains more or less constant), then
density fluctuations have a HZ spectrum when
the equal-time power spectrum of $\phi$ has a form
\begin{equation}
k^3 P_\phi = B^2 \ ,
\end{equation}
with $B \approx 10^{-5}$. If in addition at horizon crossing
we have $\phi_h \approx \delta_h$, then we need\footnote{This may be seen
as an independent assumption, or justified using
Einstein's equations. The perturbed Friedmann equations in the comoving
longitudinal gauge
imply that $k^2 \phi \approx a^2 \rho \delta$ and, using the
Friedmann equation, $\dot{a}^2\approx \rho a^2$,
and eq.~(\ref{firstcrossing}), we obtain $\phi\approx (k_h/k)^2\delta$.
Hence $\phi_h \approx \delta_h$. }:
\begin{equation}\label{keyassumption}
k_h^3 \delta_h^2 \approx B^2 \approx 10^{-10}\ ,
\end{equation}
which is sometimes used as the definition of the HZ spectrum.
Since $\phi_h \approx \delta_h$, this expression is true in any gauge
because the $\delta_h$ defined in the various gauges are all
proportional to each other.

If we identify these fluctuations with thermal fluctuations
about to leave the horizon, then, using eq.~(\ref{amplT}),
this is equivalent to
\begin{equation}
 {T^2 \over \rho^2}{d \rho \over d T} k_h^3(T) \approx B^2  \ .
\label{ktoT}
\end{equation}
(where again we have neglected factors of order 1).
eq.~(\ref{rhoeq}) then leads to
\begin{equation}\label{condition}
{\gamma\over l_T^{\gamma-4}} k_h^3 T^{1-\gamma} \approx B^2 \ .
\end{equation}
Taking the time derivative of the above expression leads to
\begin{equation}\label{dcondt}
3 {\dot{k_h} \over k_h} + (1-\gamma) {\dot{T} \over T} = 0 \ .
\end{equation}
Since $\dot{k_h}>0$ is the condition for modes to be leaving the horizon, it
follows from eq.~(\ref{dcondt}) that
\begin{equation}\label{thecond}
(1-\gamma) {\dot{T} \over T} < 0 \ .
\end{equation}
Hence scale invariance requires that $\gamma<1$ or $\gamma>1$ depending
on whether the Universe is getting colder or hotter. We recall that $\gamma<1$ is equivalent
to saying that the thermal coherence length $\lambda$ {\it increases} with the temperature
(see eq.~(\ref{corleng})), or that
the fractional energy fluctuations increase with the temperature.
This is quite anomalous and we shall rule it out explicitly
in the next section. Hence we are left with warming universes
as a possibility for HZ fluctuations of thermal origin.

These results may be expressed more quantitatively by noting that
Eqs.~(\ref{khor}) and~(\ref{condition}) lead to
the condition for scale invariance:
\begin{equation}\label{sicond}
\gamma=1+3\mu    \ ,
\end{equation}
or more generally the expression for the tilt
\begin{equation}
n=4+{1-\gamma\over \mu} \ .
\end{equation}

In our argument so far we have abstained from using Einstein's
equations. Instead, we have treated eq~(\ref{keyassumption}) as
an independent assumption and also {\it assumed} that the gravitational potential
would freeze outside the horizon. However we can go further if we are prepared to use
Friedmann equations:
\begin{equation} \label{friedman}
{\dot{a} \over a} \propto \sqrt{\rho} \  \ {\rm and} \ \ \rho
\propto {1 \over a^{3(1+w)}} \ ,
\end{equation}
(where $w$ is the equation of state), and if we
parameterize the dependence of the speed of light on the
temperature as:
\begin{equation}\label{cvsT}
c \propto T^{\alpha} \ .
\end{equation}
Then from eqns.~(\ref{firstcrossing}), (\ref{khor}),
(\ref{friedman}) and (\ref{cvsT}) the condition for scale
invariance reduces to:
\begin{equation}\label{scalecon}
{\gamma\over 1-3\alpha}=2{1+w\over 1-w}
\end{equation}
or, alternatively, to:
\begin{equation}\label{scalecon1}
\alpha={(2-\gamma)+(2+\gamma)w\over 6 (1+w)}  \ .
\end{equation}

\section{A no-go theorem for cooling universes}\label{nogo}
If at all times the universe has been cooling ($\dot T<0$), then
we arrive at a forbidding set of conditions for a
scale-invariant spectrum. In this case $k_h$ must decrease with the temperature ($\mu<0$).
As already noted, from
(\ref{thecond}) it then follows that this requires
\begin{equation}\label{gammaconstraint}
\gamma<1
\end{equation}
This condition is a strict inequality: $\gamma=1$
leads to either $n=4$ (if the modes are indeed being pushed out of the horizon),
or to $n=0$ (if the Hubble length stagnates).

If $\gamma<1$ the fractional amplitude of
thermal fluctuations anomalously {\it increases} with the temperature.
As equation (\ref{corleng}) shows, this also implies that the thermal correlation
length increases with the temperature.
Below we present a no-go theorem which shows that it is unlikely that this condition
is satisfied assuming that the weak energy condition is satisfied and that the
fundamental degrees of freedom are particle-like.

Consider radiation in thermal equilibrium with a certain
dispersion relation $p(E)$
which becomes the usual $p^2=E^2$ at sufficiently low energies.
The energy density is proportional to the integral:
\begin{equation}
\rho(T) \propto I(T)=\int dE \ {E \ p^2(E) \over e^{E/T}-1} \ \left|{d p \over d E} \right| \ ,
\end{equation}
where the integration is over all allowed values of energy. For convenience, let us define
$F(E)\equiv Ep^2(E)\left|{d p / d E} \right|$ and re-write $I(T)$ as:
\begin{equation}
I(T)= T \ \int {dE \over T} \ {F(E) \over e^{E/T}-1} \equiv T f(T) \ .
\end{equation}
In order to have $\gamma < 1$, $f(T)$ must be
a decreasing function of temperature at sufficiently high values of T, {i.~e.} $f'(T) < 0$ .
Let us evaluate $f'(T)$ :
\begin{equation}
f'(T) = \int {dE \over T^2} {F(E) \over e^{E/T}-1} \left[{E \over T}{e^{E/T} \over e^{E/T}-1} - 1 \right] \ .
\label{fprime}
\end{equation}
If we only consider non-negative energies, then the factor under the integral appearing in
front of the square brackets is non-negative, while the expression inside the brackets is a
non-negative, monotonically increasing function of $E/T$. Hence, $f'(T) > 0$ for all $T$.
Thus, the inequality (\ref{gammaconstraint}) cannot be satisfied.

The above proof is quite general and is valid for all models that would aim to
achieve $\gamma < 1$ by modifying the dispersion relations without altering the
statistical properties of the gas. In particular, this proof implies that the modified
dispersion relations considered in Refs.~\cite{moddisp,ncvsl,ncinfl}) could not result in
a HZ spectrum.

The no-go argument assumed that the thermalized radiation was made of
particles obeying Bose-Einstein statistics. A radically different
statistics is likely to be needed in order to obtain the desired
exponent in the Stefan-Boltzmann relation. Thus the only way we
see of bypassing this no-go theorem is to allow for non-particle
like degrees of freedom. We conclude this section by suggesting a
scenario which may make use of this possibility.

\subsection{Saturating temperature} An example of a
system in which it is possible to have $\gamma<1$ is
a gas of strings at temperatures close to the so-called Hagedorn temperature,
$T_H$ \cite{hagedorn}. In a gas of strings, the number of degenerate
states increases exponentially with energy \cite{weinberg} and the canonical
partition function diverges for all $T>T_H$. This does not necessarily mean that
temperature higher than $T_H$ are unphysical. In fact, all physical quantities, such as
energy density and specific heat are actually finite at $T\ge T_H$ \cite{vafa}.
In \cite{atick} it was suggested that $T_H$ corresponds to a phase transition, somewhat
analogous to the deconfining transition in QCD. At temperatures close to $T_H$, the canonical
ensemble description of string gases becomes invalid due to increasingly large energy
fluctuations \cite{mitchell}. One must use the microcanonical ensemble instead, which is well-defined
only if all spatial dimensions were compactified \cite{vafa}\footnote{In Ref.~\cite{vafa} it is further
suggested that in this picture one needs a mechanism which would later make three of the spatial
dimensions sufficiently large for us to live in.}.

At least within the canonical ensemble formalism, $T_H$ can be interpreted as the limiting
temperature of the gas -- as energy is increased, the temperature remains constant.
In the language of eq.~(\ref{rhoeq}) this corresponds to $\gamma \rightarrow 0$, in agreement with
the constraint (\ref{gammaconstraint}). A straightforward examination of eq.~(\ref{scalecon}) with
$\gamma=0$ shows that scale-invariance can be satisfied if either $w=-1$,
as in inflation, or if $\alpha=1/3$, as in VSL models.

String-driven inflationary models, making use of the existence of a limiting temperature,
have been considered in the late $1980$'s \cite{aharon,turok88}. More recently, in Ref.~\cite{freese},
it was proposed that winding modes of open strings on D-branes above the Hagedorn phase transition
can provide the negative pressure necessary to drive inflation. In particular, it was suggested
that one could achieve a period of exponential inflation (with $w$=-1) if not all
transverse dimensions supported winding modes \cite{freese}.

It would be very interesting to investigate if thermal fluctuations could indeed be
viable candidates for structure formation in these models and the degree of
fine-tuning it would involve.

The other possibility, that of a VSL theory with $\alpha=1/3$, is currently lacking a specific
model realization.

\section{Warming universes}
\label{warming} But it could be that at the early stage when the modes
are being pushed out of the horizon the universe is getting
hotter. Such is the case of thermal radiation with $\gamma
(1+w)<0$. If $\gamma>0$, denser radiation means hotter radiation;
however if $w<-1$ the universe gets denser (and so hotter) as it
expands. Alternatively we could have $w>-1$, so that the universe
gets less dense as it expands; but then if $\gamma<0$ this
translates into a higher temperature.

Another possibility is a stage of radiation injection, either from particle/antiparticle
annihilation, from false vacuum decay, or from a cosmological
constant discharge.

Yet another possibility is the Phoenix universe of Lemaitre \cite{lem},
where modes would be pushed outside the horizon with temperature increasing
during the contracting phase.

If the universe gets hotter in time, we need $k_h$ to increase with time, and with
temperature, so that $\mu>0$.
A necessary condition for scale invariance is then
\begin{equation}
\gamma>1 \ ,
\end{equation}
bypassig the no-go theorem in the previous section.
Again, from eqns.~(\ref{firstcrossing}), (\ref{khor}),
(\ref{friedman}) and (\ref{cvsT}) we obtain
\begin{equation}\label{horcon}
\mu=\gamma \left[ {1 \over 2} - {1 \over {3(1+w)}} \right] - \alpha>0
\end{equation}
or equivalently
\begin{equation}\label{horcond}
{\alpha\over \gamma} <{1+3w\over 6(1+w)} \ .
\end{equation}
Conditions (\ref{scalecon}) and (\ref{scalecon1}) still apply.

We now consider particular solutions to these conditions.

\subsection{A phantom phase}
``Phantom'' matter \cite{Caldwell:1999ew} exhibits an equation
of state with $w<-1$, and it may constitute
the dark energy of the universe. It has also been conjectured
that normal radiation at high temperatures could behave like
phantom matter \cite{ncinfl}. For these models there is a critical
density, $\rho_c$, such that $w>-1$ for $\rho<\rho_c$, while
for $\rho>\rho_c$ one has $w<-1$. If the Universe
starts off with $\rho>\rho_c$ and expanding then
$a\propto (-t)^{2\over 3(1+w)}$. As the universe expands it gets
denser and hotter. Eventually a phase transition brings it
to the sub-Planckian regime.

In such a scenario there is hyper-inflation, so the modes
are pushed out of the horizon without a VSL. However, in order
for density fluctuations to have a scale invariant spectrum,
the condition in eq. (\ref{scalecon}) must be satisfied. Since $\gamma>1$
and $w<-1$, we find that
\begin{equation}
\alpha > {1 \over 2} -{1\over 3(\omega+1)} > {1 \over 2} \ .
\end{equation}
Hence, in order to obtain scale-invariance, one needs a VSL.

Regarding the spectrum's amplitude, from eq.~(\ref{condition}) one can
obtain the requirement
\begin{equation}
{\gamma\over l_T^{\gamma-4}}\lambda_1^{(1-\mu)(1-\gamma) \over
\mu} \approx B^2   \approx 10^{-10}\ ,
\end{equation}
or, using eq.~(\ref{sicond}) to express $\mu$ in terms of
$\lambda$,
\begin{equation}\label{amplconst}
\gamma \left({\lambda_1 \over
l_T}\right)^{\gamma-4} \approx B^2  \ .
\end{equation}
Thus, given a value of $\gamma$, eq.~(\ref{amplconst}) constrains
the ratio of the two characteristic length scales $l_T$ and
$\lambda_1$. Within any specific model providing a relation between the scale
factor $a$ and the temperature $T$ it should be possible to express $\lambda_1$
in terms of $l_T$. Consequently, these two length scales should not be considered
as independent.

We may expect one of the models based upon the formalism
presented in \cite{ncvsl} to satisfy these conditions, placing
further restrictions upon deformations of dispersion relations.
The spectrum's amplitude in this case will be fixed by the ratio
between the non-commutative length scale (related to $l_T$) and the
Planck length (related to $\lambda_1$).

\subsection{A bouncing universe}
\label{bouncing}
In the bouncing model a closed universe goes through a series of
cycles starting with a Big Bang and expansion, followed by
re-contraction and a Big Crunch. Every time the universe
approaches the crunch it bounces into a new bang, and if
entropy increases at the bounce the new cycle lasts longer
(i.e. expands to a larger and colder turnaround point) in each
new cycle. Although the idea goes back to Lemaitre~\cite{lem},
it has often been forgotten and revived a few times,
e.g. more recently in \cite{durrer}.

It is questionable that this model dispenses with fine-tuning,
specially when dealing with the homogeneity and entropy problems.
It may be that just too much junk (entropy) is left over from previous cycles.
This can usually be avoided combining the bouncing universe
with inflation, such as in its ekpyrotic incarnation
\cite{ekp}.

Another possibility is that the speed of light decreases at the
bounce, so that all the relevant scales we see today were
sub-horizon modes at some point deep in the radiation epoch in the
contracting phase. Hence
homogeneity (and absence of pervading coalescing black holes)
could have been established on all the relevant scales for the
current cycle. But more importantly in this scenario thermal
fluctuations could become the Harrisson-Zeldovich spectrum of
initial conditions which we observed in our current cycle. We now
illustrate how this could happen.

As long as $w>-1$ (and $\gamma>0$) the universe heats up in time
during the contracting phase. Hence a
necessary condition for scale invariance is that $\gamma>1$,
evading the no-go theorem presented in Section~\ref{nogo}. Modes
must leave the horizon as the universe contracts, and as
eq.~(\ref{horcon}) shows this can be achieved without a VSL if
$w>-1/3$, i.e. the universe undergoes the necessary accelerated
contraction when the strong energy condition {\it is} satisfied. A
general condition for scale invariance is then eq.~(\ref{scalecon}).
We find the notable result that standard radiation ($w=1/3$, $\gamma=4$, $\alpha=0$)
satisfies the condition for scale invariance. Hence,
as long as the scales we can observe today were sub-horizon
deep in the radiation epoch during the contracting phase of the previous cycle,
thermal fluctuations in standard radiation lead to a Harrisson-Zeldovich
spectrum of initial fluctuations in the new phase\footnote{This assumes certain
mode matching at the bounce. In general, the picture will be much more
complicated. This issue is discussed in subsection \ref{modematching}}.

A problem arises with the amplitude, which may be remedied if there is a drop in
the speed of light at the bounce. For $\gamma=4$ the scale $l_T$ drops
out of the problem (cf. eq.~(\ref{rhoeq})). Also for $w=1/3$ and $\alpha=0$
we have $k_h\propto T$, so $\mu=1$ and the scale $\lambda_1$ disappears from
eq.~(\ref{khor}). Hence the ratio of these parameters disappears from the
expression for the amplitude, as can be seen directly from Eqn.~(\ref{amplconst});
the amplitude becomes then a combination of numerical factors of order 1
clearly in contradiction with observations.  This can also be illustrated
using the $10$ Mpc comoving scale as the normalization point, as in
Peebles' example \cite{peebles}.
Rephrasing Peebles argument in terms of this model, we know
that in the case of a perfectly symmetric bounce the right normalization $\sigma_{10}$
can be obtained if the 10 Mpc comoving scale leaves the horizon in the contracting phase
when the universe was at $10^{11}$ Gev. However, if the relation between temperature
and time on either side of the bounce is symmetric, this comoving
scale would leave the horizon much earlier in the contracting phase, when
the universe was much colder and therefore the fractional fluctuations were
much larger.

\begin{figure}
\vskip 0.5 truecm
\epsfxsize= 0.95\hsize\epsfbox{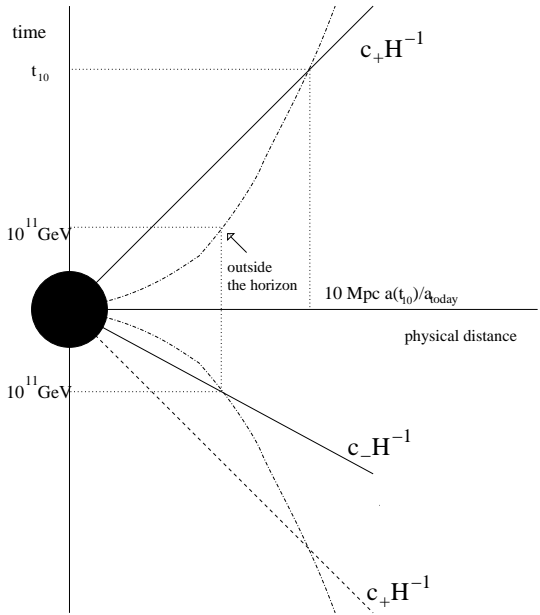}
\vskip 0.5 truecm
\caption{\label{fig1} The normalization problem in bouncing universes and a way to solve it.
In the case of a perfectly symmetrical bounce, density perturbations on $10$ Mpc scale exit
the horizon during the contracting phase at temperatures much lower than $10^{11}$ GeV. The problem
could be fixed if the speed of light in the contracting phase ($c_-$) had a much bigger constant value
than it has today ({\it i.~e. $c_- \gg c_+$}).}
\end{figure}

The problem with the normalization mentioned above, could be resolved if the
bounce was asymmetrical, allowing for the 10 Mpc comoving scale to leave
the horizon when the universe was indeed at $10^{11}$ Gev. One possible scenario is when
the speed of light is constant during the previous
($c = c_-$) and current ($c= c_+$) cycles, but $c_-\gg c_+$, i.e.
the speed of light decreases at the bounce. One can make additional simplifying
assumptions just to illustrate the point. Let us assume that the fundamental
constants $G$ and $\hbar$ also change at the bounce in such a way so as
to preserve the relation between the temperature and time (this can be achieved if
$\hbar^3 c^5/G$ is the same before and after the bounce). Hence the relation
between time and temperature is symmetrical around the bounce, but for
$c_-/c_+\ll 1$ the $10$ Mpc comoving scale would leave the horizon in the contracting phase
much later and at a higher temperature, as illustrated in Fig.~\ref{fig1}.

Using $t/(1{\rm sec})\approx (T/(1{\rm Mev}))^{-2}$,
we find that $t_\star= 10^{-28}$ sec before the bounce the temperature is
$10^{11}$ Gev. The comoving horizon at this time is 10 Mpc across
if
\begin{equation}
c_-t_\star=10{\rm Mpc} {a_\star\over a_0}
\end{equation}
Since ${a_\star/ a_0}=T_{CMB}/{10^{11}{\rm Gev}}$ (where $T_{CMB}\approx 2.7$ K)
we get
\begin{equation}\label{c-c+}
{c_-\over c_+}
\approx 10^{21}
\end{equation}
Note that unlike other VSL arguments, which lead to lower bounds on
$c_-/c_+$, this argument leads to an identity: the spectrum amplitude
results directly from a given value of $c_-/c_+$.
It may be possible to obtain a similar effect by another way of making the
bounce asymmetrical. For instance if $G$ and $\hbar$ vary differently
at the bounce a different constraint is obtained.

Certainly, the validity of the above discussion
depends upon the matching of growing and decaying modes during the
bounce dynamics - something which is unknown in the absence of a specific
model for the bounce. We conclude with a discussion of the two main
uncertainties faced by this model.

\subsubsection{Mode matching at the bounce}
\label{modematching} Bounce dynamics will in general mix the modes
in either phase\cite{vern,finelli}, a matter which may upset our
previous arguments. In the contracting phase we have
\begin{equation}
\phi=A_-\phi_0+{B_-\over |t|}
\end{equation}
and in the expanding phase
\begin{equation}
\phi=A_+\phi_0+{B_+\over t}
\end{equation}
where $\phi$ is the gravitational potential and $t=0$ at the bounce. Matching these modes
depends on the bounce dynamics, and everything we have said
previously assumed that $B_-$ mode was not excited and that the $A_-$ mode
was matched onto the $A_+$ mode.

If the $B_-$ mode is excited, but then matches on to the $B_+$
mode, then the earlier discussion still applies. However if the
$B_-$ mode matches onto the $A_+$ mode then the conclusions in
this paper have to be revised: in addition to the HZ spectrum
predicted here, there would be a very red component ($n=-3$).
Another pathological case is when $A_-$ and $B_+$ are excited,
while $B_- = A_+ =0$.

We defer to a future publication a more careful study of this
situation. It certainly depends on how the bounce is actually
produced. Mode matching in pre-big-bang\cite{pbb} and
ekpyrotic\cite{ekp} cosmologies has been discussed in considerable
detail in Refs.~\cite{vern,finelli}

\subsubsection{Entropy production at the bounce}
Another matter which may change the spectrum and normalization
of the fluctuations is entropy production at the bounce (which
almost certainly occurs).
Indeed for normal radiation, {\it i.~e.} with $\gamma=4$, the amplitude
of thermal density fluctuations on a given scale $L$ can be expressed in terms of
the entropy contained in a sphere of radius $L$ (eq.~(15.26)
of \cite{peebles}):
\begin{equation}
\delta_L^2 = {16\over 3S} \ .
\end{equation}
Thus, if entropy is produced at different rates on different scales
the spectrum of fluctuations could be modified. In addition this
issue is bound to interfere with any normalization condition for the
amplitude.

\section{Conclusions}
\label{summary}
We have studied the possibility of thermal fluctuations providing seeds for
currently observed cosmic structures. Assuming thermal equilibrium,
we have identified the necessary conditions under which these fluctuations are Gaussian
and scale-invariant. We have shown that Gaussianity constraints are automatically satisfied
in thermal equilibrium. The situation with the scale-invariance of the
power spectrum is more problematic. In order to have a HZ spectrum
in a universe that cools in time
one needs significant modifications to the Stefan-Boltzmann law,
namely, $\rho \propto T^\gamma$ with $\gamma <1$. We have shown that this
condition is unlikely to be satisfied for radiation
comprised of Bose-Einstein particles. If we are prepared to assume
that the radiation is not made of particles, then naturally a number
of possibilities could open up.
If over a certain period the Universe was warming up with time,
then the condition for scale invariance becomes $\gamma >1$ and
can, in principle, be achieved with a gas of Bose-Einstein
particles with appropriate dispersion relations.

An approach different from ours was taken in Ref.~\cite{labini}. There, authors have
considered thermal fluctuations in a system of particles with an attractive
short range potential and a repulsive $1/r^2$ potential at large scales.
They have shown that the resulting power spectrum of density fluctuations
is scale-invariant on sufficiently large scales (small $k$s). We find
this direction of thought interesting and deserving further development.

To conclude, we have discussed an alternative, ``thermal'', way of
obtaining scale-invariant initial spectrum of density
fluctuations. Other ways to match the observations without
inflationary quantum fluctuations have been previously discussed,
{\it e.~g.} active seed models \cite{defect} with \cite{avel} or
without VSL \cite{dmatch}. One may argue that some of these models
are less ``natural'' than inflation. However, they make the point
that the recent observational victories in modern cosmology are a
success of the Harrisson-Zeldovich spectrum plus gravitational
instability rather than a  ``proof'' of the inflationary origin of
the initial fluctuations.

\acknowledgements
We are grateful to Ruth Durrer for pointing out several errors in an earlier
draft of the manuscript, and to Pedro Ferreira for sharing
his ideas on the subject and for bringing Ref.~\cite{labini} to our attention.
We also thank Andrew Liddle for clarifying comments on the definition
of HZ spectra and choice of gauge.

\end{document}